\documentclass[aps,pra,preprint,groupedaddress,amssymb]{revtex4-1}
\usepackage{latexsym}
\usepackage{amsbsy}
\usepackage{amsmath}
\usepackage{graphicx}
\usepackage{color}

\begin{document}

\title{Interlayer Dzyaloshinskii-Moriya interactions.}

\author{Elena Y. Vedmedenko}
\affiliation{Department of Physics, University of Hamburg, D-20355 Hamburg, Germany}
\author{Patricia Riego}
\affiliation{CIC nanoGUNE, E-20018 San Sebastian, Spain}
\affiliation{Departamento de Fisica de la Materia Condensada, Universidad del Pais Vasco, UPV/EHU, E-48080 Bilbao, Spain}
\author{Jon A. Arregi}
\author{Andreas Berger}
\affiliation{CIC nanoGUNE, E-20018 San Sebastian, Spain}

\begin{abstract}
Interfacial Dzyaloshinkii-Moriya interaction defines a rotational sense for the magnetization of two-dimensional films and can be used to create chiral magnetic structures like spin-spirals and skyrmions in those films. Here we show by means of atomistic calculations that in heterostructures magnetic layers can be additionally coupled by an interlayer Dzyaloshinskii-Moriya interaction across a spacer. We quantify this interaction in the framework of the L\'evy-Fert model for trilayers consisting of two ferromagnets separated by a non-magnetic spacer and show that interlayer Dzyaloshinkii-Moriya interaction yields non-trivial three-dimensional spiral states across the entire trilayer, which evolve within as well between the planes and, hence, combine intra- and inter-plane chiralities. This analysis opens new perspectives for three-dimensional tailoring of the magnetization chirality in magnetic multilayers.
\end{abstract}

\maketitle

The magnetic Dzyaloshinskii-Moriya interaction (DMI) arises in systems with bulk inversion asymmetry  \cite{Fert1990,Bogdanov1994a}. Without bulk inversion asymmetry, the DMI arises at interfaces only and couples two magnetic sites both sitting within a surface layer \cite{Fert1990,Fert1992}. This interaction appeared to be a very important property of interfacial systems, because it is responsible for the unique rotational sense of magnetization and can be used to create topological objects like magnetic skyrmions and chiral domain walls \cite{Yu2010,Sampaio2013,Romming2015,Woo2016}, that are attractive candidates for data storage, transfer and processing  \cite{Lin2013,Kiselev2011,Fert2013}. DMI corresponds to an antisymmetric part of the exchange tensor and is described by a vector quantity $\vec D$. Orientation and strength of $\vec D$ can be estimated using the Moriya symmetry rules \cite{Moriya1960}, the L\'evy and Fert model \cite{Fert1992} or first-principles calculations \cite{AnatomyDMI,Bluegel,Udvardi}. The Moriya procedure has been created for localized magnetic systems and takes into account two magnetic sites coupled by a Hubbard-type Hamiltonian. The L\'evy and Fert model involves an additional third site mediating the DMI via conducting electrons and is more appropriate for itinerant systems. In most cases symmetry rules as well as three-sites model give correct orientation of $\vec D$. Both of the models have, though, their limitations. The two-sites procedure can often predict only an easy plane, rather than an exact direction  of $\vec D$ \cite{Crepieux}. DMI from three-sites model applied to systems of low-symmetry like spin chains at interfaces might differ in some cases from ab-initio results \cite{Bluegel}. Nontheless, it is broadly accepted that for ultrathin films the L\'evy-Fert model provides sound basis for studies on the spin ordering at the interfaces, because majority of experimentally 4d/3d, 5d/3d interfaces or their alloys belong to the class of itinerant systems. Its additional advantage is a clear definition of $\vec D$ in systems with large and complicated unit cell or in disordered systems, which are difficult to treat from the first principles.

The typical strength of the DMI at interfaces lies between 0.1 and 2 meV per atomic bond  \cite{Heide2008,Dupe2015,Hagemeister2015,Rozsa2015}, which corresponds to the thermal energy of several tens of Kelvin. To enhance the DMI in view of room temperature applications, ..NM1/FM/NM2.. multilayers have been proposed \cite{Fert2016,Dupe2015,Kang:SciRep2018}. In these multilayers a strong intralayer (within FM layers) and interlayer (between FM layers) exchange or Rudermann-Kittel-Kasuya-Yoshida (RKKY) coupling was considered, while the DMI appeared within the NM1/FM or FM/NM2 interfaces only. The RKKY coupling ensures identical magnetic behavior of all FM layers. Hence, if the DMI at all interfaces have identical direction, they can be added to enhance the total DMI and a complete stack behaves like an entity. Interlayer DMI across a NM spacer has up to now not been addressed despite the fact that NM atoms or impurities within the NM layer might play a role via the "third site" coupling of magnetic layers. The reason why the interlayer DMI has not been considered yet is, probably, the strong decrease of the DMI with the distance between interacting and mediating sites as shown in \cite{AnatomyDMI,Landeros2016} and cancellation of DM vectors for certain symmetries. If, however, the DMI across a spacer exists, different physical scenarios might develop: - if the interlayer DMI supports the RKKY coupling it might further enhance the effective DMI and approach the technological breakthrough; - if the interlayer DMI competes with the RKKY coupling some unexpected phenomena like intrinsic separation of columnar skyrmions, bias-effects or three-dimensional frustration might emerge.

In this study, we employ analytical and atomistic Monte-Carlo (MC) calculations to investigate the DMI coupling between two FM layers across a non-magnetic spacer. We find that the interlayer DMI exist for many microscopic geometries.  Despite the weakness of interlayer DMI per atomic bond, it can induce chiral coupling between FM layers, because the total interlayer DM energy creates a sizable energy barrier between macroscopic configurations with different chirality. This coupling is not trivial and seeks to create a three-dimensional spin spiral across the complete system. Therefore, competition between the interlayer DMI and other energy contributions opens a new class of frustrated magnetic systems and can be used to enhance the effective interfacial DMI or to create three-dimensional complex magnetic structures.

\begin{figure}[tb]
\includegraphics[width=\textwidth]{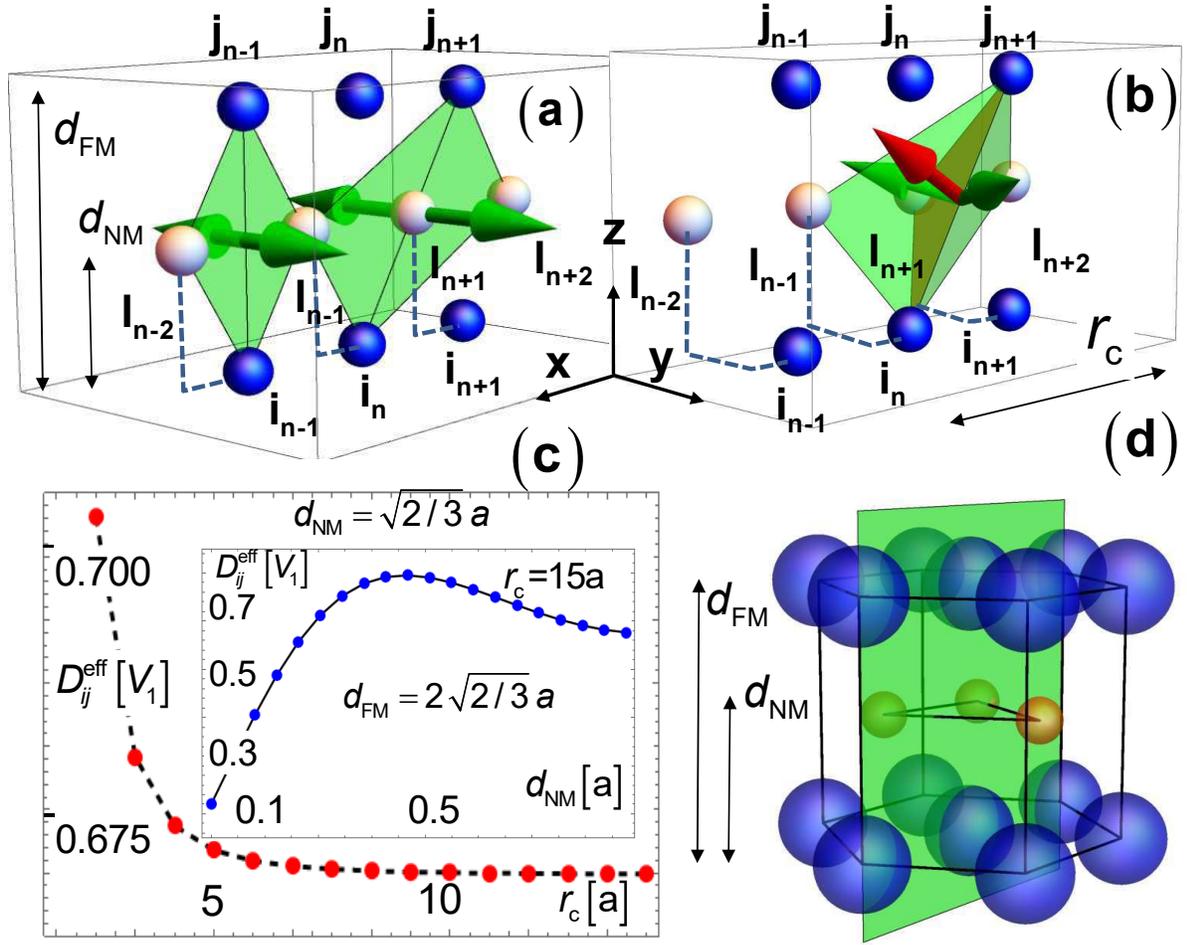}
\caption{Microscopic interlayer Dzyaloshinskii-Moriya vectors in a one-dimensional trilayer. \textbf{(a,b)} Contributions to the interlayer DM vectors ${\vec D}_{ij}$ due to mediating NM sites to the right and to the left from the bond. In \textbf{(a)} magnetic and non-magnetic atoms lie in the same plane and ${\vec D}_{ij}^{left}={-\vec D}_{ij}^{right}$ leading to $D_{ij}^{\rm eff}=0$. In \textbf{(b)} NM atoms are shifted along $y$ axis corresponding to a crossection of an $hcp$ stacking along a dense-packed direction shown in \textbf{(d)}. Because of the shift, $D_{ij}^{\rm eff}\neq 0$.  \textbf{(c)} Effective interlayer DM vector  $\vec D_{ij}^{\rm eff}$ for a chain with geometry \textbf{(b)} in dependence of the number $n$ of mediating $l$ atoms ($r_{\rm c}=a\cdot n$). The inset gives function  $D_{ij}^{\rm eff}=f(d_{\rm NM})$ for a given $d_{\rm FM}$ and $r_{\rm c}=15a$; \textbf{(d)} Three-dimensional analog of the one-dimensional structure in (b).}
\label{fig:F1}
\end{figure}

In our calculations magnetic layers are represented by monolayers of Heisenberg spins $\vec S_i$ at atomic positions $\vec R_i$ (see Fig. 1). The two effective FM monolayers are separated by a NM metallic layer. The distance between the bottom FM layer and the NM layer is $d_{\rm NM}$. This NM layer is assumed to contain a certain distribution of impurity atoms at positions  $\vec R_l$, which can each mediate a DMI between any two spins according to the three-site model by L\'evy and Fert \cite{Fert1992}:
\begin{equation}\label{Eq:Levy}
 {\vec D_{ijl}}({\vec R_{li}},{\vec R_{lj}},{R_{ij}}) = \\ - {V_1}\frac{{\sin \left( {{k_F}({R_{li}} + {R_{lj}} + {R_{ij}}) + (\pi /10){Z_d}} \right)\left( {{{\vec R}_{li}} \cdot {{\vec R}_{lj}}} \right)\left( {{{\vec R}_{li}} \times {{\vec R}_{lj}}} \right)}}{{|{R_{li}}{|^3}|{R_{lj}}{|^3}{R_{ij}}}},
\end{equation}
where ${\vec R}_{li}$, ${\vec R}_{lj}$ are the distance vectors from the impurity $l$ to corresponding FM atom sites $i$ and $j$, and $\vec R_{ij}$ the distance vector between these FM sites. The parameter $V_1= \frac{{135\pi }}{{32}}\frac{{{\lambda _d}{\Gamma ^2}}}{{E_F^2k_F^3}}\sin \left( {\frac{\pi }{{10}}{Z_d}} \right)$ refers to the material specific quantity defining the DMI strength. Hereby, $k_F$ and $E_F$ are the Fermi wave vector and energy respectively, ${\lambda _d}$  spin-orbit coupling parameter, $\Gamma$  interaction parameter between the localized spins and the spins of conduction electrons and $Z_d$ the number of d-electrons. This sum is oscillating and, hence, non-trivial. According to \cite{Fert1992} $V_1$ is strong not only for heavy transition metals like Pt(5d) and Pd (4d), but also for 3d Ni, Co, and Fe-impurities. An effective DM vector of a given $ij$ atomic pair can be described by a sum over all impurities $l$ \cite{Landeros2016,Fert1992}:

\begin{equation}\label{Eq:Landeros}
{\vec D}_{ij}^{\rm eff} = \sum\limits_{l} {{{\vec D}_{ijl}}({{\vec R}_{li}},{{\vec R}_{lj}},{R_{ij}})}
\end{equation}

First we calculate interlayer DM vectors for simplest possible one-dimensional case with fully symmetric positions of atoms and impurities shown in Fig. 1a. In this case, individual ${\vec D_{ijl}}$ are symmetric with respect to the $n$-th $ij$ bond and add up to zero (arrows in Fig. 1a) according to the Eq. \ref{Eq:Landeros}. On the other hand, different geometries of mediating atoms are possible. A simplest 1d configuration, which is different from that of Fig. 1a is shown in Fig. 1b. This geometry correspond to the ($1\bar 100$) plane of an $hcp$ structure (Fig. 1d) and impurities are shifted in the $-y$ direction. Because of this shift, an additional $\vec D_{ijl}$ emerge (e.g. $\vec D_{i_{n+1}j_{n+1}l_{n}}$ in addition to $\vec D_{i_{n+1}j_{n+1}l_{n+1}}$ and $\vec D_{i_{n+1}j_{n+1}l_{n+2}}$). All vectors are not compensated anymore. Hence, an interlayer $\vec D_{ij}^{\rm eff}$ becomes non-zero.  Fig. 1c shows strength of ${\vec D}_{ij}^{\rm eff}$ in Fig. 1b as a function of the number of considered $l$-atoms denoted by the distance $r_{\rm c}=n\cdot a$ with $a$ distance between $l$ atoms. It shows that $|{\vec D}_{ij}^{\rm eff}|$ is maximal ($\approx 0.7 V_1$) if nearest neighboring impurities only are considered and decreases non-significantly (to $\approx 0.675 V_1$) if further nearest neighbors are taken into account. The dependence $|{\vec D}_{ij}^{\rm eff}|=f(d_{\rm NM})$  (see inset of Fig. 1c) shows the variation of $|{\vec D}_{ij}^{\rm eff}|$ with the position of NM layer. Calculated $\vec D_{ij}^{\rm eff}$ for a complete $hcp$ unit cell are shown in Fig. 2a,b. Apart from an ideal $hcp$ stacking, the $\vec D_{ij}^{\rm eff}$ exist for all orthorombic lattices, but vanish for ideal trilayers with cubic $(001)$, $(110)$ and $(111)$ surfaces. However, as soon as $d_{NM}\ne (d_{FM}-d_{NM})$ the interlayer DM vectors become finite. The latter situation is a standard case at interfaces, particularly when two different FM layers are used. The interlayer DM vectors are particularly strong at the grain boundaries, where the crystal structure departs from its ideal stacking and impurities take on-top positions (see Supplementary information).

\begin{figure}
\includegraphics[width=0.8\textwidth]{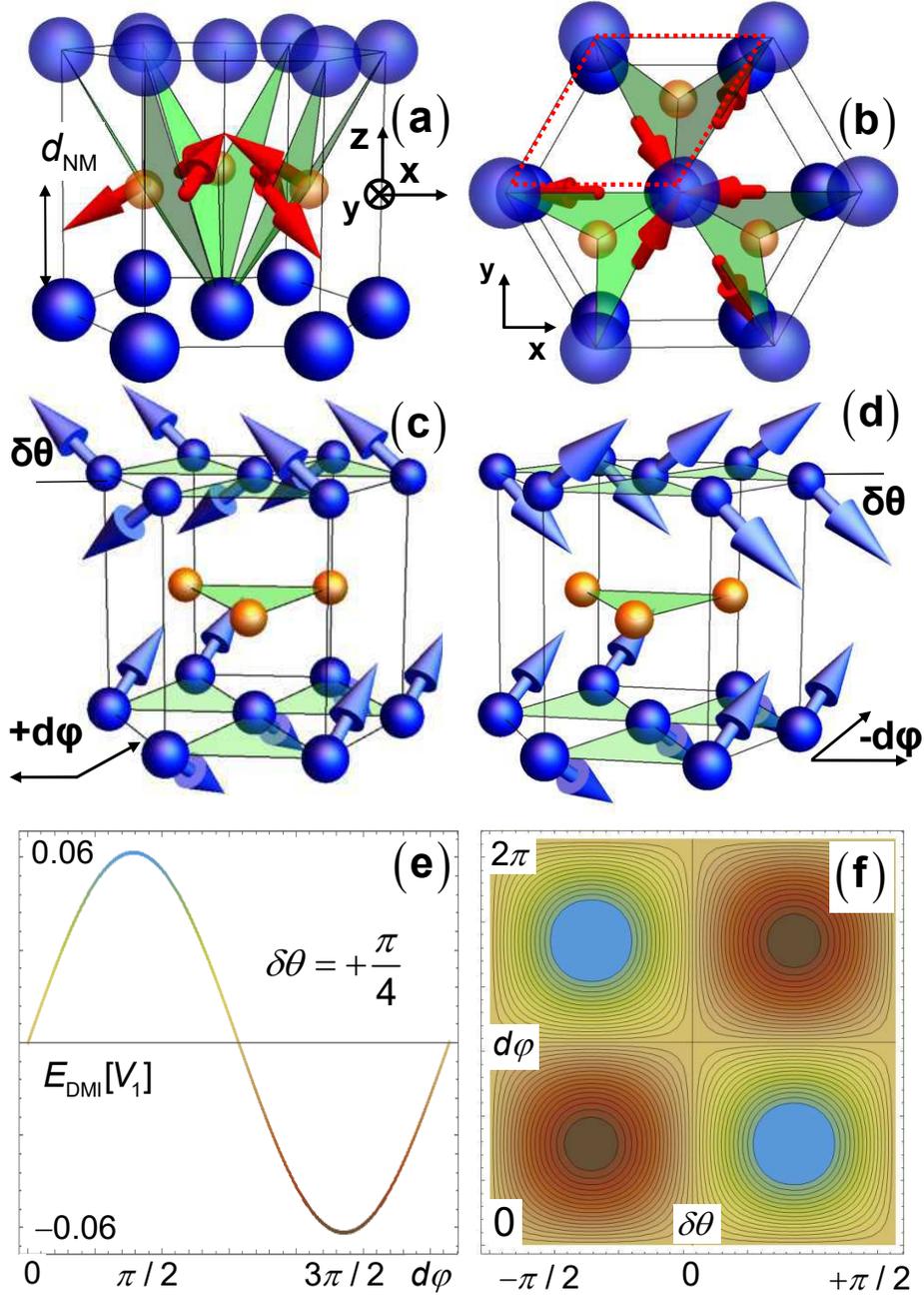}
		\caption{ Interlayer Dzyaloshinskii-Moriya energy in a FM/NM/FM trilayer. Orientations of DM vectors calculated on the basis of 3-sites-model for a trilayer with an $hcp$ structure: (a) side-view; (b) top-view. Green triangles define the $ijl$ planes, dark spheres show magnetic moments, light balls correspond to mediating atoms; (c,d) Analytically minimized azimuthal rotations of the top FM layer with respect to the bottom FM layer for an up-down (c) and down-up (d) magnetization oscillations in the top layer. The DM energy is minimized by the $d\phi=+\pi/2$ rotation in (c), while by $d\phi=-\pi/2$ in (d); (e,f) Energy landscape in the $d\phi,\delta\theta$ phase space as shown in panels (c), (d).}
		\label{fig:F2}
\end{figure}

To understand how microscopic ${\vec D}_{ij}^{\rm eff}$ can generate net interlayer DMI, we calculate the total DM energy for an FM/NM/FM trilayer with $hcp$ stacking (see Fig. 2a,b) analytically:
\begin{equation}\label{Eq:Mixed}
   E_{\rm DM} = \sum\limits_{ij} \vec D_{ij}^{\rm eff}\cdot{{\left( {{{\vec S}_i} \times {{\vec S}_j}} \right)}}
\end{equation}
First, we analyse $E_{\rm DM}$ for the case of FM1 and FM2 being each in a perfectly aligned ferromagnetic state, but allowed to have any orientation with respect one to another. Because the spin cross-product is identical and constant for all pairs, the DM energy per site is
\begin{equation}\label{Eq:FM}
E_{\rm DM}^i ={{\left( {{{\vec S}_i} \times {{\vec S}_j}} \right)}}\cdot\sum_{j}{\vec D}_{ij}^{\rm eff}.
\end{equation}
While individual $\vec D_{ij}^{\rm eff}$ are non-vanishing, their sum $\sum_{j}{\vec D}_{ij}^{\rm eff}$ vanishes, because all $\vec D_{ij}^{\rm eff}$ for a site $i$ in a hexagon cancel out as can be seen in  Fig. 2a,b. Hence, perfectly aligned FM layers do not interact via interlayer DMI despite non-zero $\vec D_{ij}^{\rm eff}$.
However, if magnetization configurations in each of the layers exhibit deviations from perfect ferromagnetic alignment, the spin cross-product cannot be taken out of the sum and indices of the two vectors cannot be separated. Hence, Eq.~\ref{Eq:Mixed} cannot be converted into the simple form of Eq.~\ref{Eq:FM}. Instead, spin and distance variables become mixed and even if $\sum_{j}{\vec D}_{ij}^{\rm eff}$ vanishes, the total $E_{\rm DMI}^i$ might become non-zero. Hence, while perfectly aligned ferromagnetic layers do not exhibit DMI-induced interactions via a spacer, this might be possible if deviations from perfect collinear alignment within the magnetic layers are present.

To illustrate this we first conduct a simplified analytical minimization of the total DM energy calculated on the basis of Eq.~\ref{Eq:Mixed} with respect to noncollinear magnetic states. For this, we assume that azimuthal spin angles $\phi_i$ within each FM layer are identical, while $\phi_i^{FM1}$ and $\phi_i^{FM2}$ can take any values. We furthermore assume row-wise up-down deviations $\pm \delta\theta$ of spins from $\theta=\pi/2$ as visualized in Fig. 2c,d. The magnitude of these polar deviations is identical in both layers. However, for a given sequence of $\pm \delta\theta$ in the bottom layer we have distinguished between up-down (see Fig. 2c) and down-up (see Fig. 2d) sequences in the top layer. By applying these constraints we simplify the problem and reduce the system to having two state variables ($\delta\theta$ and $d\phi$) only. Fig. 2e,f show analytically calculated interlayer $E_{\rm DM}$ as a function of $\phi_i^{FM1}$, $\phi_i^{FM2}$, and $\delta\theta$ for these two sequences. The DM energy is minimized by $\delta\theta=\pi/4$ and $d\phi=+ \pi/2$ for the up-down sequence (see Fig. 2e), but by $\delta\theta=\pi/3$ and $d\phi=- \pi/2$ for the down-up sequence (Fig. 2e). Corresponding configurations are depicted in Fig. 2c,d. These calculations permit to make three important conclusions: (i) Perfectly aligned FM layers are decoupled from the point of view of interlayer DMI; (ii) Magnetic noncollinearity promotes interlayer DM coupling. Hereby, net DM energy depends on the relative orientation of the net magnetizations within the two layers and, thus, constitutes a net interlayer coupling of net ferromagnetic states; (iii) The sign of the interlayer chirality depends on the phase of the magnetization oscillations within the layers around some net orientation. Hence, by only considering the interlayer DMI, and despite constrains of solution space, we find magnetization states, that lower the DM energy, while at the same time exhibiting a net ferromagnetic moment in each of the two FM layers.

\begin{figure}
\includegraphics[width=0.8\textwidth]{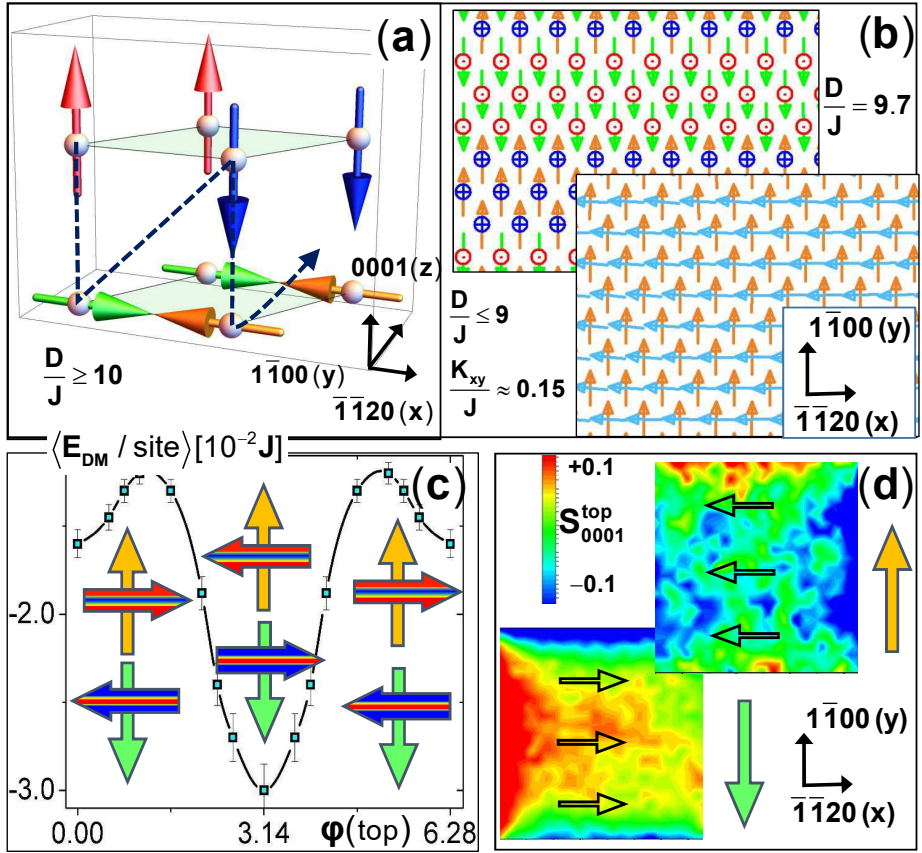}
		\caption{Ground states of trilayers with interlayer DM interaction from MC simulations. (a) Bottom panel: top view of a portion of stable MC low-temperature ($k_BT=0.001 J_{ij}^{\rm intra}$) configuration for $|{\vec D}_{ij}^{\rm eff}|/J_{ij}^{\rm intra}=30$ and atomic structure of Fig. 1d; Top panel: Three-dimensional representation of one period of the magnetic structure in bottom panel with the same color scale (blue and red correspond to opposite out-of-plane , while green and cyan to opposite in-plane magnetization). A cycloidal spin spiral across the complete system is formed.  (b) MC evolution of the mean interlayer DMI per magnetic bond for two different temperatures, when simulations started with perfect ferromagnetic alignment of two layers and $90^{\circ}$ rotation of the top layer with respect to the bottom one. Symbols correspond to the numerical data with standard deviation bars, while solid lines are exponential fits of the numerical data; (c) Mean interlayer DM energy achieved in simulations described in (b) for different initial states. Initial magnetization orientation is visualized by arrows (the bottom layer is always horizontal). Red/blue arrow colors correspond to positive/negative out-of-plane components evolved in course of the simulations; (d) Stable MC low-temperature ($k_BT=0.001 J_{ij}^{\rm intra}$) configurations for $|{\vec D}_{ij}^{\rm eff}|/J_{ij}^{\rm intra}=0.15$, $K_{\rm x}^{\rm bot}=0.1J_{ij}^{\rm intra}$ and $K_{\rm xy}^{\rm top}=0.1J_{ij}^{\rm intra}$ for the structure of Fig. 1d. Arrows show net in-plane , while colors the out-of-plane magnetic contrast.}
		\label{fig:F3}
\end{figure}

For the general understanding of the role of interlayer DMI for magnetic structuring, we used Monte-Carlo (MC) simulations. Samples with lateral dimensions of up to $30a\times 30a$ with periodic and open boundaries have been considered. We use an $hcp$ stacking presented in Fig. 1d. The magnetic Hamiltonian
\begin{equation}
H=-\sum_{\langle ij\rangle}J_{ij}^{\rm intra} ({\vec S_{i}}\cdot{\vec S_{j}})+K_{xy}^i\sum\limits_{i}({S}_{i}^{z})^2
-\sum_{\langle ij\rangle}\vec D_{ij}^{\rm eff}({\vec S_{i}\times\vec S_{j}})
\end{equation}
includes ferromagnetic intralayer Heisenberg exchange $J_{ij}^{\rm intra}$, interlayer DM interactions $\vec D_{ij}^{\rm eff}$ as well as easy-plane $K_{xy}^i$ anisotropy accounting for shape effects. First we set to zero all contributions except of $\vec D_{ij}^{\rm  eff}=V_1$. The corresponding ground state of the energy of $\langle E_{\rm DM}\rangle\approx -0.82V_1$/atom is depicted in Fig. 3a. It is an antiferromagnetic row-wise ordering in both layers. Hereby, magnetization in one of the layers is perpendicular to the film plane, while the other rotates by $\pi/2$; i.e., is planar. Whilst both layers are individually antiferromagnetic, collinear and do not show any chirality, the total state demonstrates clockwise cycloidal spin spiral across the both layers (see black path in Fig. 3a) and counterclockwise cycloidal rotation between the layers. Hence, unique chirality between the layers and a unique modulation within the layers define a unique magnetic chirality across the layers. This non-trivial structure is different from a typically one-dimensional bulk spin spiral with perfect ferromagnetic order in each plane. In the next step we add ferromagnetic intralayer exchange interaction. For all $\vec D_{ij}^{\rm eff}/J_{ij}^{\rm intra}>10$ the ground state remains the same as the DMI dominate. In the narrow region $9\leq\vec D_{ij}^{\rm eff}/J_{ij}^{\rm intra}\leq 10$  the antiparallel rows broaden (see Fig. 3b, top inset) and reach the sample size. Vertical magnetization component in both layers, however, show wave-like deviations from corresponding in-plane or out-of-plane magnetization. The DM energy drops to $\langle E_{\rm DM}\rangle\approx 0.025 J_{ij}^{\rm intra}$/atom for $\vec D_{ij}^{\rm eff}\approx J_{ij}^{\rm intra}$ and $K_{xy}=0$. If both layers have perfect FM order the interlayer DM energy vanish. Hence, in contrast to standard bulk DMI, a necessary condition for formation of a chiral rotation between two planes is magnetic noncollinearity.

It appears interesting to investigate whether the interlayer DMI can govern chirality of the large class of in-plane multilayers. To force magnetization into film plane, we add an easy-plane anisotropy and use parameters typical for Co based alloys \cite{Eyrich:JAP2012}: $J_{ij}^{\rm intra}=10$ meV per atomic bond, $V_1=0.15 J_{ij}^{\rm intra}$, and shape anisotropy $K_{xy}^{\rm top}=0.15  J_{ij}^{\rm intra}$. An additional small uniaxial in-plane anisotropy in the bottom plane $K_{\rm[\bar{1}\bar{1}20]}^{\rm bot}$ gives a preferential in-plane orientation to make the study more transparent. Typical stable magnetic states are analyzed in Fig. 3b-d. Net in-plane magnetization in both layers is close to unity, while vertical magnetization varies between $S_z=+0.1$ and $-0.1$. Portion of MC ground state obtained after slow cooling from a random initial configuration is shown in bottom inset of Fig. 3b. It shows global helicoidal $\pi/2$ rotation of the top layer with respect to the bottom one with unique rotational sense. Hence, the MC simulations show that even weak interlayer DM can define global chirality of in-plane multilayers.

To clarify the physical grounds of this chirality we have performed additional analysis. We started MC simulations using perfect in-plane ferromagnetic states in each layer making an angle $d\phi$ with respect one to another as initial configuration and relaxed these states at low temperature ($kT< D_{ij}^{\rm eff}$) until the DM energy started to oscillate around its minimal mean value. Obtained mean energies $\langle E_{\rm DM}\rangle$ for different $d\phi$ are plotted in Fig. 3c. Because of the low temperature $\langle d\phi\rangle$ remained close to its initial value. At the same time, magnetization acquired small out-of-plane modulation. The arrow-like insets in panel Fig. 3c show initial magnetization orientation, while colors denote the out-of-plane component of magnetization for open boundary conditions. One global and one local energy minima have been found. The global energy minimum corresponds to the clockwise $\pi/2$ rotation of the top layer with respect to the bottom one, while the local energy minimum to an anticlockwise rotation. Hereby, there are always specific modulation of the vertical component of magnetization in the top layer, which inherits the chirality along vertical $[0001]$ axis of a pure DM ground state shown in Fig. 3a. For instance, if bottom layer is magnetized in $+y$ direction (orange arrows) the top layer acquires down-up-down (red-blue-red) modulation; if bottom layer is oriented along $-y$ axis, the modulation is reversed. Magnitude and spatial distribution of magnetization is shown in Fig. 3d.  Hence, the interlayer DMI for in-plane magnetic bilayers with dominating exchange interactions leads to a unique interlayer chirality combined with specific and associated $S_z$ modulation: the change of the sign in chirality requires sign change in the phase of out-of-plane modulations and vice-versa. The interlayer coupling emerges due to these modulations, while fully collinear layers do not interact via interlayer DMI.

The strength of the interlayer DM energy in planar bilayers is, according to our calculations, of the order of $10^{-2}J_{ij}$ per atom, which is small. An important question arises how can this weak coupling compete with other interactions or determine ordered magnetic configurations? The answer to this query lies in the collective nature of this interaction, similar to the superparamagnetism. The interlayer noncollinearity is intimately connected with the intralayer state determined by strong energy contributions and the temperature. While the interlayer DM cannot compete with these contributions at the atomic scale, it can define the energy barrier between two global configuration with different relative magnetization orientations of individual layers (see Fig. 3c). This energy barrier is given by the total $\langle E_{\rm DM}^{\rm inter}\rangle$. Thus, the $\langle E_{\rm DM}^{\rm inter}\rangle$ becomes increasingly important with increasing sample size. In our specific case it reaches the value of $\approx J_{ij}$ in a sample consisting of $2\times 10^3$ spins only and, therefore, can co-define the equilibrium state. We expect an important impact of the interlayer DMI on the three-dimensional chirality of magnetic multilayers with dominating three-sites mechanism of DMI in systems like Fe/Cr/Fe \cite{Gruenberg:PRL1986}, Co/Cu/Co \cite{Parkin:PRL1991}, Co/Ru/Co \cite{Parkin:PRL1990}, Fe/Mo/Fe \cite{Berger:PRL1992} and related Co or Fe alloys like Co/Pt/CoFeB or CoFeB/Ru/CoFeB \cite{Wiese:APL2004,Fert2016}. The strength of the interlayer DMI can be controlled by the thickness and atomic structure of the NM layer.

The main conclusion of this investigation is that additionally to the now well explored interfacial DMI, magnetic layers can be strongly coupled by means of a so-far neglected interlayer DM interaction across a mediating layer. The driving mechanism of the interlayer DM coupling is the formation of the global chiral structure across magnetic multilayer in all directions (in-plane and out-of-plane), which is strongly different from a one-dimensional bulk spin spiral. It might be particularly important in systems dominated by ferromagnetic exchange. Hereby, a given up-down sequence of the microscopic deviations from the collinearity defines the sign of the interlayer chirality. The microscopic characteristics of this interaction depend on the lattice geometry of FM/NM/FM stacks and on the strength of the spin-orbit coupling parameter $V_1$. This finding opens completely new perspectives for enhancement and/or manipulation of the total DM interaction and magnetic structuring in magnetic multilayers.




\end{document}